\documentclass[prb,reprint,notitlepage,amsmath,amsfonts,amssymb,floatfix]{revtex4-2}
\setcitestyle{super,open={},close={}}
\usepackage[T1]{fontenc}
\usepackage[utf8]{inputenc}
\usepackage{graphicx}
\usepackage{bm}
\usepackage{color}
\usepackage{subcaption}
\usepackage{physics}
\usepackage{import}
\usepackage{gensymb}
\usepackage{booktabs}
\usepackage[hidelinks]{hyperref}

\newcommand{\Eh}{E_{\mathrm h}}
\newcommand{\mEh}{\mathrm{m}E_{\mathrm h}}
\graphicspath{{./}}

\begin{document}
\title{Multi-symmetry projected coupled cluster theory with triples corrections}

\author{Ruiheng Song}
\affiliation{Department of Chemistry, Rice University, Houston, TX 77005-1892}

\author{Thomas M. Henderson}
\affiliation{Department of Chemistry, Rice University, Houston, TX 77005-1892}
\affiliation{Department of Physics and Astronomy, Rice University, Houston, TX 77005-1892}

\author{Gustavo E. Scuseria}
\email{guscus@rice.edu}
\affiliation{Department of Chemistry, Rice University, Houston, TX 77005-1892}
\affiliation{Department of Physics and Astronomy, Rice University, Houston, TX 77005-1892}
\date{\today}

\begin{abstract}
Symmetry projection and coupled cluster theory provide complementary descriptions of static and dynamic correlation. We introduce two approximate perturbative triples corrections to spin- and point-group-projected coupled cluster with single and double excitations (PCCSD). The coherent $(T)$ correction combines triples intermediates over the projection grid, whereas $+(T)$ evaluates the conventional triples expression using PCCSD amplitudes. Both recover conventional $(T)$ in the absence of projection. We assess these corrections for N$_2$ and C$_2$ dissociation, $3d$ transition-metal monoxides, and Cr$_2$. Projection improves the singles-and-doubles description and distinguishes selected low-lying symmetry sectors. For the six-monoxide benchmark subset, coherent $(T)$ reduces the PCCSD mean absolute error from $13.4$ to $3.0~\mEh$, achieving accuracy comparable to conventional CCSD(T). Both triples corrections restore the short-bond minimum missing from the PCCSD curves of Cr$_2$, with coherent $(T)$ giving a well depth closer to extrapolated DMRG. In contrast, $+(T)$ provides more balanced N$_2$ and C$_2$ dissociation curves. These results demonstrate the benefits of perturbative triples in PCC, while their system-dependent performance highlights the need for a more systematic treatment.
\end{abstract}
\maketitle

\section{Introduction}
Coupled cluster (CC) theory provides a systematic and accurate description of electron correlation for weakly-correlated molecules.\cite{Cizek1966CoupledCluster,Bartlett2007CoupledCluster,Paldus1999CriticalCC} The singles-and-doubles approximation, CCSD,\cite{Purvis1982CCSD} supplemented by a perturbative triples correction, CCSD(T),\cite{Raghavachari1989Triples,Watts1993OpenShellTriples,Scuseria1991CCSDTGradients} is widely regarded as the ``gold standard'' for such systems. Applications of coupled-cluster methods to large systems have been enabled by local correlation and other cost-reduction strategies. \cite{Riplinger2013DLPNO,Riplinger2013Triples,Li2002CIM,Li2006CIM,Guo2018DLPNOT1,Nagy17JCP,Nagy19JCTC,GyeviNagy21JCTC,Song2026RPALNOCC,Ye2024PeriodicLNOCC,Yang2026ISDFEmbedding,Schutski2017TensorCC,Parrish2019RankReducedCC,Lesiuk2022RankReducedCCSD,Hohenstein2022RankReducedCCIII,Lesiuk2020CCSDTTensor,Jiang2023TensorHypercontraction}
However, CCSD(T) becomes less impressive when several configurations become important, as in stretched multiple bonds and many transition-metal compounds. In such cases, strong static correlation can require higher-order excitations or alternative treatments that account explicitly for the multireference character of the wave function.

A common approach to strong correlation is to describe the important configurations explicitly within an active space. Complete active space self-consistent field (CASSCF) theory\cite{Roos1980CASSCF,Siegbahn1981CASSCF} provides such a reference, with dynamic correlation subsequently recovered through methods such as CASPT2\cite{Andersson1990CASPT2,Andersson1992CASPT2} or NEVPT2.\cite{Angeli2001NEVPT,Angeli2001SCNEVPT2,Angeli2002NEVPT2}
Multireference CC offers another route to combining a multiconfigurational reference with an exponential treatment of correlation.\cite{Jeziorski1981MRCC,Lyakh2012MultireferenceCC,Evangelista2010MkMRCCT,Evangelista2011icMRCC,Evangelista2012SequentialMRCC,Evangelista2018MRCCPerspective,Hanauer2011icMRCC,Hanauer2012icMRCCT} These developments motivate compact wave functions that describe both static and dynamic correlation across different bonding regimes. However, multireference CC is far from straightforward.

Allowing the reference determinant to break symmetry can improve the description of bond dissociation,\cite{JimenezHoyos2011GHF} but at the cost of losing the correct symmetry quantum numbers---the familiar symmetry dilemma.\cite{Lowdin1955SymmetryProjection,Lykos1963HartreeFock} A subsequent truncated coupled-cluster treatment does not, in general, restore the desired spin or spatial symmetry. Symmetry projection recovers the correct symmetries and provides a different construction of a multiconfigurational state. Following the ideas of L\"owdin,\cite{Lowdin1955SymmetryProjection} projected Hartree--Fock (PHF) restores the symmetries of a broken-symmetry determinant and optimizes that determinant in the presence of projection.\cite{Mayer1980SpinProjectedEHF,Schmid2005Hubbard,JimenezHoyos2012PHF} The projected state is a linear combination of nonorthogonal determinants whose projection weights are determined by symmetry, while any required component-mixing coefficients are optimized variationally. The broader projected quasiparticle framework also permits particle-number, spin, spatial, and complex-conjugation symmetry restoration.\cite{Schmid2004ProjectedHFB,Scuseria2011ProjectedQuasiparticles} These approaches capture important static correlations without specifying an active space, but additional dynamic correlation is needed for quantitative energies.

Several methods have been developed to recover this missing correlation by incorporating particle-hole excitations on the broken-symmetry reference, including spin-projected configuration interaction,\cite{Tsuchimochi2016ProjectedCI} spin-extended coupled electron pair approximations,\cite{Tsuchimochi2017SpinExtendedCEPA} and projected perturbation theories.\cite{Tsuchimochi2019ProjectedPT2,Tsuchimochi2021ProjectedPT2} Other approaches introduce correlators acting directly on the projected reference, including killing operators,\cite{Henderson2019GeminalCI,Henderson2020AGP} Jastrow operators\cite{Dutta2020GeminalReplacement} and non-linear correlators.\cite{Khamoshi2021NonlinearAGP} Polynomial representations of spin-projected unrestricted and generalized Hartree--Fock states have also established connections with cluster expansions and enabled combinations with CC.\cite{Qiu2017PolynomialPHF,Henderson2017SGHFPolynomial}

Direct symmetry projection of a CC wave function has been developed in several related formulations. Duguet introduced symmetry-broken and restored CC for angular momentum,\cite{Duguet2015SymmetryRestoredCC} subsequently extended to particle number.\cite{Duguet2017ParticleNumber} Qiu and co-workers formulated projected CC (PCC) for molecular and model Hamiltonians using disentangled cluster operators\cite{Qiu2017PCC,Qiu2019NumberProjectedCC} and demonstrated the importance of optimizing amplitudes in the presence of projection.\cite{Qiu2018OptimizedPCC} An alternative spin-extended CC approach approximates the exponential through a truncated Taylor expansion.\cite{Tsuchimochi2018SpinExtendedCC} The power series formulation combines this strategy with disentangled cluster operators,\cite{Song2022PowerSeriesPCC} and subsequent work incorporates complex-conjugation and spin-flip projection.\cite{Song2024ComplexProjection}

An important remaining question is how to include triple excitations in PCCSD. Projection generates correlations beyond the excitation rank of the underlying cluster operator, but does not eliminate the need for additional dynamic correlation. Related work on approximate coupled-pair methods likewise highlights the importance of incorporating three-body clusters when treating strong correlation.\cite{Magoulas2022ThreeBodyClusters} Here, we compare two approximate perturbative corrections that recover conventional $(T)$ when projection is absent. 

We examine spin projection of unrestricted and generalized references, combined with point-group projection where needed to distinguish spatial symmetry sectors. N$_2$ and C$_2$ dissociation test bond recoupling and competing states. Transition-metal monoxides and Cr$_2$, for which accurate reference calculations are available,\cite{Hait2019TransitionMetals,Larsson2022ChromiumDimer} provide more demanding tests of the balance between static and dynamic correlation. Our aim is to assess the performance of symmetry projection and the two triples prescriptions across these regimes.

\section{Theory}
In this section, we briefly review symmetry projection and the basic framework of projected coupled cluster (PCC). More detailed discussions of the disentangled cluster formalism and its power series approximation can be found in Refs.~\citenum{Qiu2017PCC,Qiu2018OptimizedPCC,Song2022PowerSeriesPCC}. Throughout, $i,j,k,l,\ldots$, $a,b,c,d,\ldots$, and $p,q,r,s,\ldots$ denote occupied, virtual, and general spin orbitals, respectively.

\subsection{Projection operators}
For continuous symmetries, projection operators can be expressed as integrals over symmetry-generated transformations. For discrete symmetries, the integrals are replaced by sums. Here, we consider spin and point-group symmetry projection. The spin projection operators are
\begin{subequations}
\begin{align}
P^s_{mk} &= \ket{s,m}\bra{s,k} \nonumber\\
&= \frac{2s+1}{8\pi^2}\int d\Omega\,D^{s*}_{mk}(\Omega)R(\Omega),\\
R(\Omega) &= e^{i\alpha S_z}e^{i\beta S_y}e^{i\gamma S_z}.
\end{align}
\label{eq:spin-proj}
\end{subequations}
Here, $\ket{s,m}$ is a spin eigenstate, $D^s_{mk}(\Omega)=\mel{s,m}{R(\Omega)}{s,k}$ is a Wigner rotation matrix element, and $\Omega=(\alpha,\beta,\gamma)$ denotes Euler angles. We use $d\Omega=d\alpha\,\sin\beta\,d\beta\,d\gamma$, with $0\leq\alpha,\gamma<2\pi$ and $0\leq\beta\leq\pi$.

The point-group projection operators are
\begin{equation}
P^j_{\lambda\kappa}=\frac{l^j}{h}\sum_{g\in G}\Gamma^{j*}_{\lambda\kappa}(g)R_g,
\label{eq:pg-proj}
\end{equation}
where $h$ is the order of the group $G$, $l^j$ is the dimension of the irreducible representation $\Gamma^j$, and $\Gamma^j_{\lambda\kappa}(g)$ is the corresponding representation matrix element for the group operation $g$, where $R_g$ is the action of $g$ on the electronic wave function.

For either symmetry, projected components can be combined as
\begin{equation}
\ket{\Psi} =
\sum_{y} c_{y}P^{Z}_{xy}\ket{\Phi}.
\label{eq:projected-state}
\end{equation}
Here, $Z$ denotes the target irreducible representation, and $x,y$ label its components. This notation includes spin projection, with $(Z,x,y)=(s,m,k)$, and point-group projection, with $(Z,x,y)=(j,\lambda,\kappa)$. 
For simultaneous projection, the labels are composite, and the operators are products of the corresponding spin and point-group operators. The coefficients $c_y$ are determined from a generalized
eigenvalue problem. This mixing removes the dependence of the projected energy on the orientation of the underlying determinant within the symmetry group.\cite{JimenezHoyos2012PHF}

\subsection{Projected Hartree--Fock}
We incorporate the mixing coefficients from both the bra and ket into the contracted operator
\begin{equation}
P = \sum_{yy'} c_y^* c_{y'} P^Z_{yy'}.
\label{eq:contracted-projector}
\end{equation}
Using the multiplication and conjugation properties of the projection operators, the projected Hartree--Fock (PHF) energy is then
\begin{equation}
E_{\mathrm{PHF}}=\min_{\ket{0}}
\frac{\mel{0}{HP}{0}}{\mel{0}{P}{0}},
\label{eq:phf}
\end{equation}
where $\ket{0}$ is a broken-symmetry determinant. The coefficients $c_y$ are optimized through the generalized eigenvalue problem described above and held fixed at their converged PHF values during the subsequent PCC calculation.

We minimize this functional using an effective-Fock formulation.\cite{JimenezHoyos2012PHF}
The effective Fock matrix is
\begin{equation}
F=F_0+\bar\rho\,F_+\,\bar\rho+\rho\,F_-\,\rho,
\label{eq:phf-fock}
\end{equation}
where $\rho_{pq}=\mel{0}{c_q^\dagger c_p}{0}$ is the one-particle density matrix of the reference and $\bar\rho=1-\rho$. The first term is the orbital gradient,
\begin{equation}
(F_0)_{pq}=\frac{\mel{0}{c_q^\dagger c_p(H-E)P}{0}}{\mel{0}{P}{0}}+\mathrm{h.c.},
\label{eq:phf-gradient}
\end{equation}
and has only occupied--virtual and virtual--occupied blocks that vanish at convergence. Here, $E$ is the PHF energy at the current determinant. The remaining terms are associated with electron attachment and removal via a generalized Koopmans construction.\cite{Day1974GeneralizedKoopmans,Smith1975ExtendedKoopmans} We have defined
\begin{subequations}
\label{eq:phf-diagonal}
\begin{align}
(F_+)_{pq}&=\frac{1}{2}\frac{\mel{0}{c_p(H-E)c_q^\dagger P}{0}}{\mel{0}{P}{0}}+\mathrm{h.c.},
\\
(F_-)_{pq}&=-\frac{1}{2}\frac{\mel{0}{c_q^\dagger(H-E)c_pP}{0}}{\mel{0}{P}{0}}+\mathrm{h.c.},
\end{align}
\end{subequations}
and these terms, due to the presence of the density matrices $\rho$ and $\bar{\rho}$, define the occupied-occupied and virtual-virtual blocks of $F$ at convergence. Diagonalization of $F$ yields the PHF orbitals and the orbital energies used in the coherent perturbative triples correction below.

\subsection{Projected coupled cluster}
The projected coupled cluster energy and amplitude equations are\cite{Qiu2017PCC}
\begin{subequations}
\begin{align}
E&=\frac{\mel{0}{HPe^T}{0}}{\mel{0}{Pe^T}{0}},\label{eq:pcc-energy}\\
R_\mu&=\mel{\mu}{(H-E)Pe^T}{0}=0.\label{eq:pcc-residual}
\end{align}
\label{eq:pcc}
\end{subequations}
Here, $\bra{\mu}$ denotes an excited determinant relative to $\ket{0}$, and the cluster operator $T$ contains single and double excitations. The cluster amplitudes are optimized in the presence of the projection operator, with the PHF reference held fixed.

The central computational task is to evaluate the Hamiltonian and overlap kernels involving a symmetry rotation and the coupled cluster wave function. We use the disentangled cluster formalism \cite{Qiu2017PCC} schematically
\begin{equation}
Re^T\ket{0}=e^W\ket{0},
\label{eq:disentangled-state}
\end{equation}
where $W$ contains a scalar term and pure excitation operators. A power series expansion gives\cite{Song2022PowerSeriesPCC}
\begin{subequations}
\begin{align}
Re^T\ket{0}&=\left(1+\bar T_R+\frac{1}{2}\bar T_R^2+\cdots\right)R\ket{0}\nonumber\\
&=\left(1+\bar T_R+\frac{1}{2}\bar T_R^2+\cdots\right)e^{U_0+U_1}\ket{0}\nonumber\\
&=e^{U_0+U_1}\left(1+\bar T_{RU}+\frac{1}{2}\bar T_{RU}^2+\cdots\right)\ket{0}.
\end{align}
\label{eq:disentangle}
\end{subequations}
The barred operators denote the successive similarity transformations of $T$, and the second line follows from Thouless' theorem.\cite{Thouless1960HFStability} After acting on the reference, the resulting state can be written as $C_0(1+C_1+C_2+\cdots)\ket{0}$, where $C_n$ creates $n$-fold excitations. 
The disentangled cluster operators are then written as
\begin{subequations}
\label{eq:ci-cc}
\begin{align}
e^{W_0}&=C_0,\\
W_1&=C_1,\\
W_2&=C_2-\frac{1}{2}C_1^2,
\end{align}
\end{subequations}
with analogous expressions at higher excitation rank. In this work, the singles contribution is treated through Thouless’ theorem, and the remaining transformed doubles exponential is expanded through second order. We retain only $W_0$, $W_1$, and $W_2$.

Unlike previous formulations,\cite{Qiu2017PCC,Song2022PowerSeriesPCC} we apply the rotation operator $R$ to the ket rather than the bra. This avoids the additional similarity transformation of the Hamiltonian generated by the rotation-dependent Thouless de-excitation operator associated with the bra. The Hamiltonian kernels are therefore evaluated directly using the original integrals and the disentangled operators. The resulting formulation is equivalent to the previous one but is more compact and better suited to the perturbative triples treatment below.

Symmetry projection introduces linear dependencies among the projected excitations. Following the treatment of redundant residual directions in earlier work,\cite{Tsuchimochi2018SpinExtendedCC,Tsuchimochi2019ECC} we solve reduced residual equations,
\begin{equation}
r_\mu=\sum_\nu S_{\mu\nu}R_\nu = 0,
\label{eq:reduced-residual}
\end{equation}
where $S_{\mu\nu}=\mel{\mu}{P}{\nu}$ is the overlap matrix between projected excited determinants. This procedure addresses redundancies in the residual equations but does not entirely eliminate the corresponding redundant components of the cluster operator.

\subsection{Perturbative triples with projection}
Once the single and double amplitudes have been obtained, we estimate the contribution of triple excitations perturbatively. The projection operator complicates a direct extension of conventional CCSD(T). We therefore introduce an approximate correction whose unprojected limit recovers the usual $(T)$ expression.

Using the disentangled representation, the PCCSD energy in Eq.~\ref{eq:pcc-energy} can be written as
\begin{equation}
    E_{\mathrm{PCCSD}} =
    \frac{\sum_g w_g e^{W_0(g)}
    \mel{0}{\bar H_{W(g)}}{0}}
    {\sum_g w_g e^{W_0(g)}}.
\end{equation}
Here, $w_g$ contains the projection and quadrature weights, and $\bar H_{W(g)}$ denotes the Hamiltonian similarity-transformed by the disentangled excitation operator.

Motivated by this kernel-weighted expression, we define an approximate triples intermediate by combining the connected triples kernels constructed from $W_2(g)$ with the same normalized weights:
\begin{equation}
  \mathcal W^{abc}_{ijk} =
  \frac{\sum_g w_g e^{W_0(g)}
  \mel{^{abc}_{ijk}}{\big(V_NW_2(g)\big)_c}{0}}
  {\sum_g w_g e^{W_0(g)}}.
\label{eq:triples-w}
\end{equation}
Here, $V_N$ is the normal-ordered two-body part of the Hamiltonian, and the subscript $c$ selects connected contributions. This expression is a proposed approximation motivated by the structure of conventional $(T)$, rather than an exact projected perturbative triples. It retains the conventional connected contraction evaluated with $W_2(g)$ and neglects contributions involving higher-rank disentangled operators.

Following the same prescription, we define the disconnected intermediate as
\begin{equation}
  \mathcal V^{abc}_{ijk} =
  \frac{\sum_g w_g e^{W_0(g)}
  \mel{^{abc}_{ijk}}{
  \big(V_NW_1(g)+F_NW_2(g)\big)_{\mathrm{disc}}}{0}}
  {\sum_g w_g e^{W_0(g)}}.
\label{eq:triples-v}
\end{equation}
Here, $F_N$ is the normal-ordered one-body part of the Hamiltonian, and the subscript $\mathrm{disc}$ selects the disconnected triples contributions. The correction is evaluated as
\begin{equation}
E_{(T)}=\frac{1}{36}\sum_{ijkabc}\frac{\big(\mathcal W^{abc}_{ijk}+\mathcal V^{abc}_{ijk}\big)^*\mathcal W^{abc}_{ijk}}{D^{abc}_{ijk}},
\label{eq:triples-energy}
\end{equation}
where the orbital-energy denominator is
\begin{equation}
D^{abc}_{ijk} =
\epsilon_i+\epsilon_j+\epsilon_k
-\epsilon_a-\epsilon_b-\epsilon_c.
\label{eq:triples-denominator}
\end{equation}
Motivated by earlier work on Brueckner coupled cluster theory\cite{Scuseria1994BDT} and the multireference character of the projected state, we construct this denominator using orbital energies obtained from the PHF effective Fock operator. We call this the coherent $(T)$ correction because the projection-grid contributions are combined at the intermediate level before the final energy contraction.

A simpler alternative is to evaluate the conventional, unprojected $(T)$ expression using the converged PCCSD amplitudes and add the resulting correction to the PCCSD energy. We denote this procedure by $+(T)$. Its denominators are obtained from the semicanonicalized Fock matrix of the broken-symmetry reference. 

In either case, the $T_1$ and $T_2$ amplitudes are those of PCCSD. It is easy to see that when there is no projection, PCCSD and PCCSD(T) reduce to the common UCCSD and UCCSD(T). When the deformed determinant is RHF, they reduce to RCCSD and RCCSD(T).

\section{Results}
\subsection{Computational details}
The PHF and PCC calculations were performed with an in-house Python/Fortran implementation \cite{Song2026PHFPCC} using PySCF\cite{Sun2020PySCF} and Libcint\cite{Sun2015Libcint} for the electronic integrals. Spin projection used 8-point Gauss--Legendre quadrature for unrestricted references and a $14\times6$ Lebedev--Gaussian grid for generalized-reference PCC.\cite{Lestrange2018VAPPHF} The PHF references were optimized separately before solving the PCC equations. For generalized references, we applied the gauge-fixing procedure described in Appendix~\ref{app:gauge-fixing} to select the underlying determinant before the PCC calculation.

SUCCSD and SGCCSD denote spin-projected CCSD based on unrestricted and generalized determinants, respectively. The labels $+\mathrm{coh}(T)$ and $+(T)$ distinguish the triples corrections, and a $C_4$ prefix indicates point-group projection. We use $C_4$ to distinguish the rotational symmetry sectors considered here; the molecular term symbols identify the target states.

PCCSD convergence is assessed using the maximum absolute reduced residual, $\max |r_\mu|$, with a threshold of $5\times10^{-5}$. We adopt this relatively loose threshold as a compromise between energy convergence and iteration count: PCCSD energy changes near this threshold are typically of order $10^{-4}~\Eh$, while tighter thresholds require substantially more iterations.

All electrons were correlated for N$_2$ and C$_2$ in cc-pVDZ. The metal oxides were treated in def2-SVP at the geometries of Ref.~\citenum{Hait2019TransitionMetals}, with the metal $1s$, $2s$, and $2p$ shells and oxygen $1s$ shell frozen. Cr$_2$ calculations used cc-pVDZ-DK, a spin-free one-electron exact two-component (X2C) Hamiltonian, \cite{Peng2012ExactDecoupling,Kutzelnigg2005Quasirelativistic} and a frozen Ne core on each atom.

\subsection{Main-group molecules}
\begin{figure}[!tbp]
\centering
\includegraphics[width=1.0\linewidth]{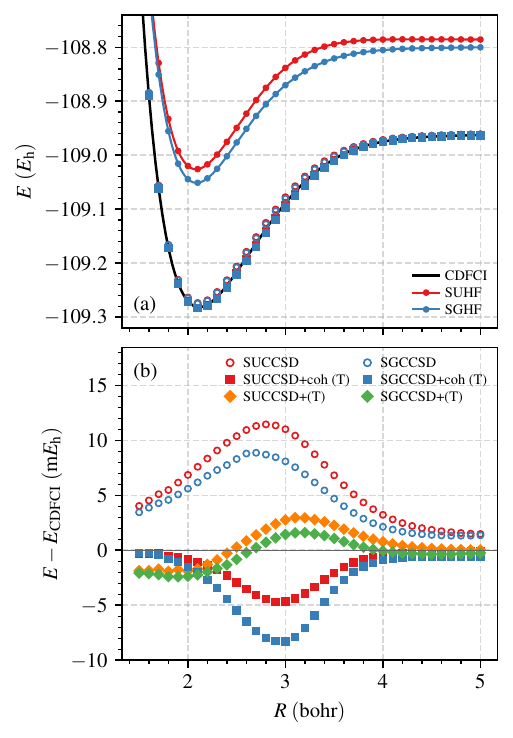}
\caption{N$_2$ dissociation in cc-pVDZ. (a) Total energies from CDFCI, PHF, PCCSD, and PCCSD$+\mathrm{coh}(T)$. (b) Errors relative to CDFCI for PCCSD and both triples corrections.}

\label{fig:n2}
\end{figure}

\begin{figure}[!tbp]
\centering
\includegraphics[width=1.0\linewidth]{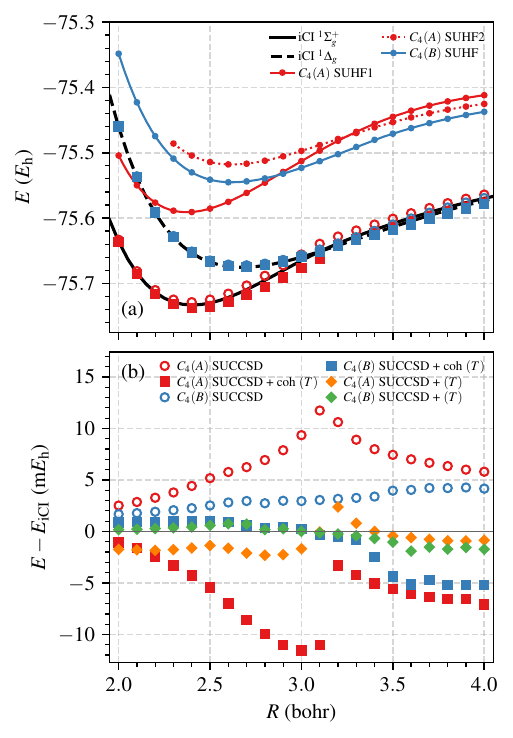}
\caption{C$_2$ dissociation in cc-pVDZ. (a) Total energies from iCI, PHF, PCCSD, and PCCSD$+\mathrm{coh}(T)$ for the $^1\Sigma_g^+$ and $^1\Delta_g$ states, targeted by $C_4(A)$ and $C_4(B)$ projection, respectively. (b) Errors relative to the corresponding iCI state for PCCSD and both triples corrections. The $C_4(A)$ PCCSD curve follows the lower of the two solution branches; triples corrections are evaluated on the selected branch.}
\label{fig:c2}
\end{figure}

We begin with the dissociation of N$_2$ and C$_2$, two familiar tests of strong correlation in stretched bonds. Because the total-energy curves are closely spaced, the upper panels of Figs.~\ref{fig:n2} and \ref{fig:c2} show PHF, PCCSD and the coherent triples correction, while the lower panels compare both triples prescriptions with their corresponding reference energies.

The N$_2$ ground-state curve has $^1\Sigma_g^+$ symmetry throughout the range considered. At the HF level, the reference becomes unrestricted beyond the Coulson--Fischer point and remains so at dissociation. In contrast, variation-after-projection (VAP) optimization yields a deformed UHF reference throughout the entire bond-length range. We further allow $S_z$ symmetry breaking to obtain a GHF reference and recover additional correlation energy upon projection. Using these optimized references, we compare spin-projected unrestricted and generalized CC with coordinate descent full configuration interaction (CDFCI) reference energies.\cite{Wang2019CDFCI} Figure~\ref{fig:n2} shows the total energies and errors for the $^1\Sigma_g^+$ ground state.

As at the PHF level, SGCCSD recovers more correlation than SUCCSD, although the deliberately introduced $S_z$ symmetry breaking yields an improvement of only a few $\mEh$. The largest PCCSD errors occur in the bond-recoupling region, reaching approximately $11.5~\mEh$ for SUCCSD and $8.9~\mEh$ for SGCCSD. The relatively large errors at equilibrium and in the recoupling region indicate the need for higher-order correlation. Both triples corrections recover the missing correlation near equilibrium. Coherent $(T)$, however, overcorrelates in the recoupling region, by about $4.7~\mEh$ for SUCCSD and $8.3~\mEh$ for SGCCSD. Thus, the better PCCSD reference does not necessarily lead to a more accurate triples-corrected energy. The simpler $+(T)$ prescription provides a more balanced curve: despite overcorrelation at short bond lengths, its errors remain within approximately $3~\mEh$ throughout the plotted range.

C$_2$ presents an additional challenge because low-lying states with the same spin quantum number but of different spatial symmetry cross along the dissociation coordinate. We target $^1\Sigma_g^+$ and $^1\Delta_g$ states with $C_4(A)$ and $C_4(B)$ projection and compare with iCI results\cite{Zhang2020ICI} in Fig.~\ref{fig:c2}. Here, we focus on spatial symmetry and only break $S^2$ symmetry to have a UHF reference.

At the PHF level, two branches of $C_4(A)$ solutions cross near 3.3 Bohr, while the $C_4(A)$ and $C_4(B)$ curves cross near $2.9$ bohr. At short bond lengths, the second $C_4(A)$ SUHF solution becomes unstable and collapses to the first. Mulliken analysis indicates that electrons preferentially occupy the $2p_x$ and $2p_y$ orbitals in the $C_4(A)$ SUHF1 solution, but the $2p_x$ and $2p_z$ orbitals in SUHF2. Complex orbitals and complex-conjugation projection may help avoid the crossing of the competing $C_4(A)$ solutions, as suggested by earlier applications,\cite{Song2024ComplexProjection} but are not considered here. 

For $C_4(A)$ PCCSD, we select the lower-energy solution from the two branches at each bond length. PCCSD reproduces the overall shape of both states and places their crossing near $3.0$ bohr, compared with approximately $3.1$ bohr in iCI. Both triples corrections shift the crossing to between $3.1$ and $3.2$ bohr. Near this crossing, the $C_4(A)$ curve switches between two PCCSD branches, producing a cusp that is amplified into a discontinuity by the triples corrections. Coherent $(T)$ strongly overcorrelates in the recoupling region of the $C_4(A)$ state and at large bond lengths in the $C_4(B)$ state. Once again, $+(T)$ gives more uniform errors, within approximately $3~\mEh$ for both states. Both molecular examples therefore favor $+(T)$ for a balanced description along the dissociation coordinate.

\subsection{Transition-metal monoxides}
The transition-metal monoxides provide a complementary test involving different spin multiplicities and spatial symmetries. We combine spin and $C_4$ projection to target the states listed
in Table~\ref{tab:mox}. Although projection onto the $C_4$ subgroup cannot distinguish all irreducible representations of the full molecular point group, it still helps reduce spatial-symmetry contamination. 
We compare with the extrapolated ASCI+PT2 reference energies of Hait and co-workers.\cite{Hait2019TransitionMetals} The conventional benchmarks labeled UCCSD and UCCSD(T) in the table use restricted open-shell Hartree--Fock (ROHF) reference orbitals, as specified in that work.

\begin{table*}[t]
\centering
\caption{Total energies ($\Eh$) and mean absolute errors (MAEs,
$\mEh$) for the $3d$ transition-metal monoxides in def2-SVP,
relative to extrapolated ASCI+PT2. Geometries and frozen-core shells follow
Ref.~\citenum{Hait2019TransitionMetals}. Conventional CC totals
were reconstructed from published bond dissociation energies
and corresponding atomic CC energies.}
\label{tab:mox}
\small
\setlength{\tabcolsep}{4pt}
\begin{tabular}{lrrrrrr}
\toprule
Species & UCCSD & $C_4$-SUCCSD & UCCSD(T) &
\shortstack{$C_4$-SUCCSD\\$+\mathrm{coh}(T)$} &
\shortstack{$C_4$-SUCCSD\\$+(T)$} & \shortstack{Extrapolated\\ASCI$+$PT2}\\
\midrule
ScO ($^2\Sigma^+$) & -834.8648 & -834.8793 & -834.8869 & -834.8906 & -834.9058 & -834.8886 \\
TiO ($^3\Delta$) & -923.5252 & -923.5429 & -923.5530 & -923.5566 & -923.5668 & -923.5547 \\
VO ($^4\Sigma^-$) & -1017.9835 & -1018.0054 & -1018.0169 & -1018.0224 & -1018.0291 & -1018.0195 \\
CrO ($^5\Pi$) & -1118.3567 & -1118.3733 & -1118.3875 & -1118.3944 & -1118.3879 & -1118.3909 \\
MnO ($^6\Sigma^+$) & -1224.8511 & -1224.8566 & -1224.8708 & -1224.8698 & -1224.8776 & -1224.8753 \\
FeO ($^5\Delta$)$^{b}$ & -1337.4430 & -1337.4460 & -1337.4663 & -1337.4648 & -1337.4710 & -1337.4710 \\
CoO ($^4\Delta$)$^{b}$ & -1456.3987 & -1456.4236 & -1456.4314 & -1456.4433 & -1456.4453 & -1456.4401 \\
NiO ($^3\Pi$)$^{a}$ & -1581.8317 & -1581.8454 & -1581.8533 & -1581.8670 & -1581.8628 & -1581.8677 \\
CuO ($^2\Pi$)$^{b}$ & -1713.9215 & -1713.9280 & -1713.9402 & -1713.9496 & -1713.9635 & -1713.9427 \\
ZnO ($^1\Sigma^+$) & -1852.7477 & -1852.7631 & -1852.7704 & -1852.7742 & -1852.7845 & -1852.7718 \\
\midrule
MAE (all ten) & 29.9 & 15.9 & 4.6 & 3.5 & 8.8 & \multicolumn{1}{c}{--} \\
MAE (six species)$^{c}$ & 28.6 & 13.4 & 2.6 & 3.0 & 9.5 & \multicolumn{1}{c}{--} \\
\bottomrule
\end{tabular}
\par\smallskip
\begin{minipage}{\linewidth}\footnotesize
$^{a}$The $^3\Pi$ assignment is inferred from the reproduced reference determinant; the symmetry of the ASCI state remains to be confirmed.
\par
$^{b}$PCCSD convergence is incomplete. The PCCSD and associated triples-corrected energies are provisional.
\par
$^{c}$Excludes FeO, CoO, and CuO because of incomplete PCCSD convergence, and NiO because of its uncertain benchmark state assignment. The same six species are used for all methods.
\end{minipage}
\end{table*}

PCCSD convergence is more difficult for FeO, CoO, and CuO: the energies change by less than $0.3~\mEh$ over the final 20 iterations, while the largest reduced residual magnitudes remain above $10^{-4}$.
Projection improves every singles-and-doubles entry in Table~\ref{tab:mox}, reducing the full-set MAE from $29.9$ to $15.9~\mEh$. PCCSD convergence remains incomplete for FeO, CoO, and CuO, and the state assignment of the NiO benchmark remains uncertain. Excluding these four species, coherent $(T)$ reduces the PCCSD MAE from $13.4$ to $3.0~\mEh$, comparable to $2.6~\mEh$ for UCCSD(T). Thus, the principal benefit of the triples correction is the recovery of correlation missing from PCCSD, bringing its accuracy close to that of conventional CCSD(T) for these equilibrium geometries.

In contrast to N$_2$ and C$_2$, $+(T)$ tends to overcorrelate the six-species subset defined in footnote~$c$ of Table~\ref{tab:mox}, with CrO being the exception. Its MAE is $9.5~\mEh$, compared with $3.0~\mEh$ for coherent $(T)$. The larger errors may partly arise from contributions associated with unwanted spatial-symmetry components, which are not explicitly removed in the unprojected $+(T)$ expression. By combining intermediates with the projection weights, coherent $(T)$ may partially suppress these contributions.

\subsection{Chromium dimer}
\begin{figure}[!tbp]
\centering
\includegraphics[width=1.0\linewidth]{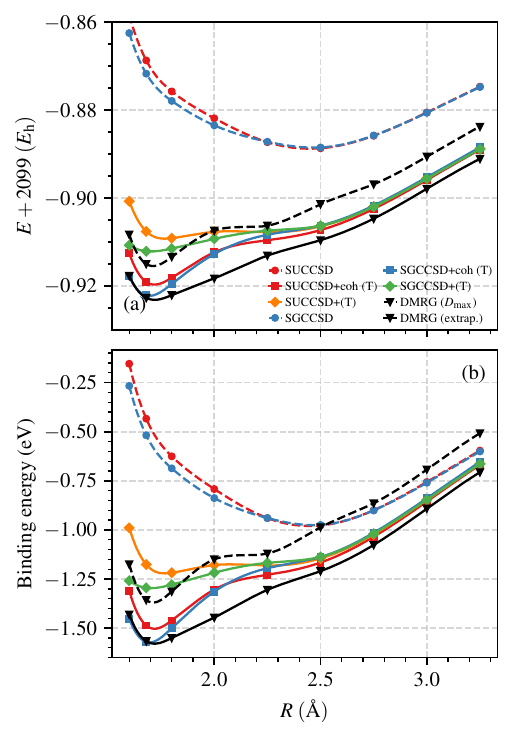}
\caption{Cr$_2$ dissociation in cc-pVDZ-DK with the spin-free X2C Hamiltonian. (a) Total energies shifted by $2099~\Eh$. (b) Binding energies. DMRG($D_{\max}$) and DMRG(extrap.) denote the largest bond dimension in the optimization and extrapolated results, respectively.}
\label{fig:cr2}
\end{figure}

Finally, we consider Cr$_2$, a longstanding challenge in quantum chemistry that requires an accurate treatment of both static and dynamic correlation. Its 12 valence electrons participate in a formal sextuple bond, giving rise to strong entanglement upon dissociation, while higher-order excitations are needed to describe correlation involving nonvalence electrons. We compare our potential energy curves with the largest-bond-dimension and extrapolated DMRG results of Larsson and co-workers.\cite{Larsson2022ChromiumDimer} Figure~\ref{fig:cr2} shows total energies and binding energies. For PCC, binding energies are referenced to the energy of each method at $R=50~\text{\AA}$, since the singlet projection does not apply separately to ground-state Cr atoms and the projection methods are not generally size-consistent. For DMRG, we use the reported atomic reference.

Both PCCSD methods miss the short-bond well and instead predict a broad minimum near $2.5$~\AA, indicating that substantial dynamic correlation remains unrecovered at the singles-and-doubles level. Adding either triples correction restores the short-bond minimum and substantially improves agreement with DMRG. 
Coherent $(T)$ gives the more accurate well depth: SGCCSD$+\mathrm{coh}(T)$ yields approximately $1.55$~eV, close to extrapolated DMRG, whereas $+(T)$ gives shallower wells of about $1.2$--$1.3$~eV. The four corrected curves approach one another beyond $2.25$~\AA \  and reproduce the shelf region. The remaining discrepancies occur mainly at intermediate bond lengths, where the triples-corrected methods still underestimate the correlation energy and more closely follow the shape of the unextrapolated DMRG curve.

Breaking $S_z$ symmetry recovers additional correlation at short bond lengths in both PCCSD and its triples-corrected variants, with the differences diminishing upon stretching. However, the N$_2$ results suggest that the close agreement between SGCCSD$+\mathrm{coh}(T)$ and extrapolated DMRG in the well depth may partly be due to error cancellation involving overcorrelation from the triples correction.

\section{Discussion and Conclusion}
The results show that symmetry projection and perturbative triples address complementary parts of the correlation problem. Projection improves the singles-and-doubles description and distinguishes selected low-lying symmetry sectors, while triples recover additional correlation near equilibrium and restore the short-bond well of Cr$_2$. The choice of triples correction is system-dependent: $+(T)$ is more balanced for N$_2$ and C$_2$, whereas coherent $(T)$ performs better for the monoxide set and the Cr$_2$ well depth. A universal non-iterative triples correction would be desirable.

Both corrections recover conventional $(T)$ in the absence of the projection operator, but the coherent construction involves additional approximations. Only the scalar, single, and double disentangled operators are retained here. The influence of projection-generated $W_3$, $W_4$, and higher terms on a triples correction remains to be determined. A more systematic treatment of the projected left state and the coupling to higher excitation manifolds would help assess the approximations used in the present construction.

Gauge invariances of the projected reference and linear dependencies among projected excitations are important practical limitations.\cite{Qiu2018OptimizedPCC,Song2022PowerSeriesPCC,Tsuchimochi2018SpinExtendedCC,Tsuchimochi2019ECC} Distinct broken-symmetry parameterizations can represent the same projected mean-field state,\cite{JimenezHoyos2020Goldstone} yet behave differently after truncating the disentangled cluster operator. Although the gauge-fixing procedure in Appendix~\ref{app:gauge-fixing} helps select the underlying reference, a satisfactory PHF state does not guarantee a stable PCCSD solution or a reliable triples correction. The reduced residual equations mitigate one aspect of the redundancy but do not remove the redundant components from the cluster amplitudes. Large or poorly determined amplitudes can therefore persist, with potentially amplified effects in $+(T)$.

The remaining overcorrelation, branch discontinuities, and convergence sensitivity show that our triples corrections are not yet uniformly reliable. Nevertheless, they support further development of PCC as a framework for combining symmetry restoration with coupled cluster correlation.

\section{Data Availability}
The data that support the findings of this study are available from the corresponding author upon reasonable request.

\section*{Acknowledgments}
This work was supported by the U.S. Department of Energy, Office of Basic Energy Sciences, under Award DE-SC0001474. G.E.S. is a Welch Foundation Chair (C-0036). We thank Guo Chen and Yanbo Fang for helpful discussions of projected Hartree--Fock theory, and Ning Zhang for providing the iCI data for C$_2$.

\appendix

\section{Gauge fixing in projected Hartree--Fock}
\label{app:gauge-fixing}
Gauge freedom allows different broken-symmetry determinants to represent the same projected state, but these determinants need not be equally suitable for subsequent correlation treatments. Here, we propose a procedure for selecting a reference within this manifold. Taking SGHF as an example, we transform the optimized PHF reference according to
\begin{equation}
\ket{\alpha,\beta,\gamma}
= e^{\alpha S_z}e^{i\beta S_y}e^{\gamma S_z}\ket{0},
\end{equation}
where $\alpha$, $\beta$, and $\gamma$ are real parameters. The nonunitary factors explore gauge directions associated with the dual representation of the Goldstone manifold.\cite{JimenezHoyos2020Goldstone} Because the $S_z$ transformations are nonunitary, they can change the norm and expectation values of the unprojected determinant.

For singlet projection, this transformation leaves the projected state unchanged. For general target spin $J$, the transformation preserves the space spanned by the projected components, provided that the mixing coefficients are transformed consistently using the inverse Wigner $D$ matrix. In practice, these coefficients can be reoptimized by solving the generalized eigenvalue problem for the transformed determinant. The optimized PHF energy is then unchanged under exact projection.

We select the transformation parameters by minimizing
\begin{equation}
\min_{\alpha,\beta,\gamma}
\left(
\frac{\mel{\alpha,\beta,\gamma}{S^2}{\alpha,\beta,\gamma}}
{\braket{\alpha,\beta,\gamma}{\alpha,\beta,\gamma}}
-J(J+1)
\right)^2,
\end{equation}
where $J$ is the target spin quantum number. This criterion brings the mean value of $S^2$ closer to its target value without changing the optimized projected energy. We also apply an analogous gauge-fixing step to bring the normalized expectation value of $S_z$ closer to its target value. These steps do not generally enforce spin purity, but are intended to select a more suitable broken-symmetry reference for subsequent PCC calculations. An analogous gauge-fixing procedure for SUHF references or discrete-symmetry projection has not been developed here.

\bibliography{ref}
\end{document}